**Local Structural Signatures of Shear Bands in Metallic Glasses via Electron Nanodiffraction**

*Huyen T. Pham, Daniel East, Chunguang Tang, Matteo Baggioli, Alessio Zaccone, Timothy C. Petersen*, Amelia C. Y. Liu**

Dr. Huyen T. Pham, Dr. Amelia C. Y. Liu
School of Physics and Astronomy, Monash University, Wellington Road, Clayton, 3800, Victoria, Australia
Email Address: huyen.pham@monash.edu, amelia.liu@monash.edu

Dr. Daniel East
Manufacturing Research Unit, CSIRO, Research Way, Clayton, 3168, Victoria, Australia Email Address: daniel.east@csiro.au

Dr. Chunguang Tang
Institute for Frontier Materials, Deakin University, Geelong, 3217, Victoria, Australia
Email Address: chunguang.tang@deakin.edu.au

Associate Professor. Matteo Baggioli
Wilczek Quantum Center, School of Physics and Astronomy, Shanghai Jiao Tong University, Dongchuan Road, Shanghai, 200240, Jiangsu, China
Shanghai Research Center for Quantum Sciences, Shanghai Jiao Tong University, Dongchuan Road, Shanghai, 201315, Jiangsu, China
Email Address: b.matteo@sjtu.edu.cn

Professor. Alessio Zaccone
Department of Physics "A. Pontremoli", University of Milan, via Celoria 16, 20133 Milan, Italy
Email Address: alessio.zaccone@unimi.it

Dr. Timothy C. Petersen
Monash Centre for Electron Microscopy, Monash University, Wellington Road, Clayton, 3800, Victoria, Australia
Email Address: timothy.petersen@monash.edu

**Keywords:** metallic glass, deformation, shear band, electron nanodiffraction, strain, centrosymmetry

**Abstract.** Structural changes in a glass due to deformation are subtle and difficult to quantify using conventional imaging and diffraction techniques. Additionally, transmission electron microscopy (TEM) sample preparation using energetic ions often causes structural modifications that are challenging to detect in disordered materials. By preparing inverted cross-sectional transmission electron microscopy lamellae of shear bands formed during bending, and employing cryogenic ion polishing to minimize preparation artefacts, we preserve the intrinsic atomic structure. Using sensitive, new parameters derived from electron nano-diffraction, we directly probe the local nano-scale structure in the plastic zone beneath surface shear steps in metallic glasses. Mapping of local centrosymmetry and strain reveals nanoscale, stripe-like regions oriented at ≈45° to the applied strain where strain has localized. These regions exhibit a high density of local atomic structures that have transformed to configurations with reduced centrosymmetry (≈1–2%), and increased magnitudes of shear and normal strain. Our results demonstrate that plastic deformation in metallic glasses arises from coordinated nanoscale structural transformations, providing direct experimental insight into a long-standing problem.

## 1. Introduction

In crystalline materials, plastic deformation is well understood through the creation and propagation of dislocations which are "tears" in the periodic order or topological defects [1]. Crystal dislocations can be detected from diffraction contrast images [2] or directly visualized using atomic-resolution scanning transmission electron microscopy (S/TEM) [3], enabling the mechanical properties of these materials to be understood and further engineered. For example, dislocations serve as effective carriers of plastic deformation and enable ductility [4]. In crystalline metals, when dislocation density increases or their mobility is restricted, dislocation strengthening occurs, resulting in enhanced material strength [5].

Metallic glasses are amorphous alloys formed by rapid cooling of the molten metal. Unlike crystalline materials, metallic glasses lack long-range order; consequently, translational symmetry and the concept of a unit cell are not applicable. As a result, their structure cannot be solved using conventional crystallography techniques. Their short-range order is composed of a spectrum of polyhedral types that are often distorted or not symmetric and they potentially possess some degree of medium-range order at longer length scales [6]. These materials show exceptionally high strength, surpassing most engineering materials, along with high elastic strain limits. However, they generally lack adequate ductility and toughness for many applications [7]. Glasses deform in a fundamentally different way to the dislocation-mediated plasticity of crystals. In fact, the fundamental mechanisms governing plasticity in amorphous solids such as glasses remain poorly understood. This is largely due to the absence of direct experimental observations in three-dimensional glasses and the limitations of phenomenological theories and simulations [8–10].

The susceptibility of metallic glasses to brittleness is due to the formation of "shear bands" when the stress exceeds their elastic limit. These shear bands are narrow regions (on the nanometer scale), which exhibit intense, localized strain and often appear as surface grooves or steps [11–13]. These areas of localized strain limit ductility and can be precursors to cracks and material failure. Under certain conditions, the presence of multiple shear bands or their propagation behaviour can enable a more complex plastic flow [14]. Significant efforts have been dedicated to both the theoretical and experimental investigation of the initiation and propagation of shear band formation [9,10,15–26]. Some consensus has been reached in recent years about the prominent role of shear-aligned Eshelby-like quadrupoles or plastic events, which align at 45° with respect to the applied strain to form the shear band [21,25,27–29]. Despite this, the nanoscale structural transformations that underpin shear band formation are still not clear. Recent studies have extensively examined deformation behavior and nanoscale mechanical response in metallic glass systems [30–32]. Other work demonstrates that nanoscale structural heterogeneity governs deformation behaviour in metallic glasses, providing key insight into the mechanisms of shear localisation and their influence on mechanical response [33]. To link local, nanoscale structural response to larger-scale strain localization and failure, S/TEM methods and in particular, sensitive electron nanodiffraction show promise.

S/TEM techniques have been widely employed to investigate structural changes in shear band areas due to their ability to image nanoscale volumes. Common imaging modes like high-resolution phase contrast bright field or high-angle annular dark field (HAADF) imaging detect changes in diffraction and mass thickness contrast, respectively. The difference between shear band areas and non-shear band areas is often not clearly discernible in these modes due to the disordered structure and very subtle structural transformations [15,34]. Other researchers using STEM-HAADF imaging observe a reduction in scattering intensity at the core of shear bands or fluctuating mass thickness contrast along the shear band, suggesting the presence of spatially localized free volume [21,25,35]. However, in projection images, the difference between reduced density and reduced thickness (for example from a crack) can only be disentangled by tomography. Diffraction-based methods using nanometer-sized converged electron beams in the S/TEM can detect local information about atomic arrangements. For example, local pair-distribution functions can reveal distinct local atomic arrangements within the shear band regions of metallic glass [36]. Local strain measurements can be quantified using electron nanodiffraction, enabling the detection of quadrupolar strain fields in metallic glasses both *in situ* and *post facto* [17,37,38].

Scanning electron nanodiffraction (SEND) (also often called 4-dimensional STEM “4D-STEM”) techniques have advanced through new direct-detector technologies and custom condenser apertures opening new opportunities for studying complex disordered materials [24,39]. In particular, insight into local structure, order and symmetry can be accessed in amorphous materials when diffraction is collected from a small scattering volume. This requires a beam tailored to the size of a nearest neighbour polyhedron and careful control of specimen thickness, typically 10–20 polyhedra, to minimize the overlap of features in transmission diffraction patterns from projected structures [40–44].

Experiments exploiting the same transmission diffraction geometry but with x-rays and colloidal glasses have shown how angular correlations and anisotropies in coherent micro-beam small angle x-ray diffraction patterns can be used to quantify different types of local structures [45–47]. Colloidal glasses are dense, disordered assemblies of microspheres that exhibit solid-like behavior and are in some senses, mesoscale analogues of metallic glasses. In particular, during an *in situ* deformation experiment, it has been shown that structural transformations within shear bands are subtle, involving coordinated changes in local symmetry and local strain of 1% or less [46].

In line with recent theoretical studies that highlighted local structural centrosymmetry as a key parameter in forecasting the elastic response of amorphous materials [48,49], this previous work on colloidal glasses evaluated the degree of local centrosymmetry from the breakdown in Friedel symmetry in the micro-beam diffraction patterns and found that indeed, this parameter was a sensitive probe of plastic deformation [46]. In a non-centrosymmetric arrangement of particles, the force imbalance on any given particle (Figure 1f), $i$ after affine distortion is $f_i = \Xi_i \gamma$ where $\gamma$ is the applied strain and $\Xi_i$ is a vector quantity, that quantifies the degree of centrosymmetry in nearest neighbours [48–50]. This force imbalance results in non-affine displacements. The shear modulus of the material, reflecting stiffness, is then composed of an affine ($G_A$) and non-affine ($G_{NA}$) contribution, with the non-affine contribution reducing the overall modulus and increasing softness $G = G_A - G_{NA}$. $\Xi_i$ can be calculated explicitly from particle coordinates, but here we employ angular symmetries in electron nanodiffraction patterns to probe this experimentally [48–50].

Shear band regions are extremely narrow (≈10 *nm* [51]) and distinguishing between deformed and undeformed regions may not be possible from S/TEM imaging methods alone [15,35]. Furthermore, reduction in local centrosymmetry seems to be a key component of structural softening prior to failure and local strain parameters can be used in well-defined geometric descriptors of plastic structural rearrangements in a glass [48]. Thus, locating a shear band with certainty is only possible when the shear band location is known either through *in situ* measurements or by using precise fiducial markers. Experimentally, *in situ* S/TEM observation of shear bands is highly challenging as they propagate rapidly over large areas, making it difficult to interrogate their structure with high resolution and sensitivity [17,18,52].

Here we perform a *post facto* analysis of the region of a shear band just below the surface shear step, where strain is anticipated to localize most strongly. Shear steps are only a few tens of nanometers in height, and these surface regions can undergo structural modification during TEM specimen preparation, due to damage induced by energetic ion beaming polishing such as local heating and knock-on damage. Because shear bands are extremely narrow, site-specific sample preparation is crucial, and preserving intrinsic atomic structure of the step in cross-sectional views is particularly difficult due to its confined geometry.

To capture the subtle structural changes induced by plastic deformation underneath shear steps, we fabricated high quality specimens using precise fiducial markers to identify the shear step. We prepared an inverted TEM lamella at the cross-sectional surface of a shear band formed by bending which induced tension and compression, respectively, at the two surfaces of the metallic glass ribbon. A thick (∼100 – 150 *nm*) cross-sectional lamella was fabricated using focused ion beam–scanning electron microscopy (FIB-SEM), followed by further argon ion polishing at cryogenic temperature (−150 °C) to the final thickness of approximately 20 *nm*. This inverted ion-beam thinning approach, combined with low-temperature, low energy and glancing angle polishing, ensured preservation of the intrinsic atomic structure underneath the shear step, while producing a thin specimen for electron nanodiffraction analysis.

We resolve local structural differences between regions underneath shear steps and far from the shear step with both sensitivity and spatial precision by analysing the angular symmetries and distortions present in electron nanodiffraction patterns. This analysis enables us to quantify and map the local degree of centrosymmetry and local structural anisotropy (strain) at the level of individual polyhedra. In a deformed metallic glass, we observed longer correlation lengths in these parameters from ∼10 *nm* thick "bands" that possessed positive normal strain (dilation), negative shear strain and a very subtle but significant reduction in local centrosymmetry. Our experimental results are in good agreement with predictions from atomistic simulations and previous diffraction-based measurements on colloidal glasses, reinforcing the validity and universality of the observed phenomena [46,47]. Here, validating simulations are performed to establish these scanning diffraction techniques at the truly *local scale*, with comparison to explicit displacements measured from a strained atomic model.

## 2. Results

### 2.1. Scanning electron nanodiffraction to map local structure

Figure 1a illustrates the geometry of SEND in a S/TEM. A quasi-parallel electron beam is focused into a small probe and scanned across the thin cross-sectional specimen in the shear step area. At each probe position, a nanodiffraction pattern is collected. Figure 1a shows how an array of nano-beam electron diffraction patterns is collected from the shear step region of the TEM specimen. The raster scan is optimised in 1.2 *Å* steps, which is substantially finer than the probe dimensions used in this study. For strain mapping with a 1.5 *mrad* convergence angle, the electron probe has a full width at half maximum (FWHM) of ~6.7 Å. For local centrosymmetry mapping using a 3.11 *mrad* convergence angle, the probe FWHM is approximately ~3.2 *Å*. These beam sizes define the spatial resolution and are optimized to increase the sensitivity to angular correlations and strain in the first-neighbour shell [40]. This spatial oversampling of 1.2 *Å* ensures projected coordination polyehdra with a diameter of 6 *Å* are sufficiently sampled. SEND patterns of the glass were obtained using an FEI Titan$^3$ 80-300 FEGTEM operating at 300 keV, equipped with Cs aberration correction for both the imaging and probe-forming lenses, and an electron counting electron microscope pixel array detector (EMPAD) [53].

Small, coherent probe beams are ideal for characterizing amorphous structure like metallic glasses, as they generate speckle diffraction patterns that directly reflect the local order within the probed volume. Because of this, speckle diffraction carries extremely rich structural information, enabling the quantification and spatial mapping of local order in glasses [40,54]. We also examine two local structural parameters that have been useful in understanding deformation in glasses, namely the local strain and the local degree of structural centrosymmetry [46,47].

We quantify local strain by analyzing the distortion of electron nanodiffraction patterns (Figure 1b). Previous research has established this method for parallel and converged beams, where continuous rings in diffraction patterns exhibit distinct elliptical distortions [17,37,38]. In our SEND configuration, highly speckled patterns with non-continuous diffraction rings demand special adaptation in the signal processing. To this end, we calculate the radial intensity centre-of-mass at 24 azimuthal angles and robustly fit a strain function (Equation 1) [17,37] to this azimuthal-binned data, as shown in Figure 1c.

$$\varepsilon(\phi) = \frac{k_0 - k_{\max}(\phi)}{k_{\max}(\phi)} = \varepsilon_{xx}\cos^2(\phi) + \varepsilon_{xy}\cos(\phi)\sin(\phi) + \varepsilon_{yy}\sin^2(\phi) \qquad (1)$$

The position of the maximum of the first ring as a function of scattering angle, $k_{max}(\phi)$ is normalized by the radius of the ideal ring in the unstrained case, $k_0$. In this case, $k_0$ was determined from the averaged nanodiffraction pattern from a non-shear band region, far from the shear step of deformed metallic glass, assumed to represent the unstrained reference, so that the mapped strain is the local deviation from the average. $\varepsilon_{xx}$ and $\varepsilon_{yy}$ are in-plane strain components along the directions parallel and perpendicular to the metallic glass surface, respectively and $\varepsilon_{xy}$ is the in-plane shear strain. Figure 1d shows the effect of changes to local strain in the nearest-neighbour environment. Pink atoms have undergone positive $\varepsilon_{xx}$ and $\varepsilon_{yy}$ and negative $\varepsilon_{xy}$.

We evaluate the degree of local centrosymmetry through Friedel microscopy, a technique that quantifies the breakdown of Friedel (or inversion) symmetry in each SEND pattern. The Fourier coefficients of the angular autocorrelation contain odd symmetries, indicating a breakdown of Friedel symmetry, or nonequivalent diffracted intensities related by inversion or a $\pi$ rotation (Figure 1e). This breakdown can arise from three sources: an imaginary component in the probe beam, Ewald-sphere curvature, and dynamical diffraction from a specimen lacking transverse inversion symmetry. For aberration-corrected electron beams and thin specimen's dynamical diffraction or multiple scattering in the presence of non-centrosymmetric atomic arrangements is the dominant effect leading to the absence of Friedel symmetry in SEND from glasses [42]. Previous work showed that local centrosymmetry can be quantified from such dynamical diffraction patterns by taking the ratio of even to odd angular symmetries in the diffraction pattern (see Experimental Section) [24,42,46]. Figure 1f shows a schematic diagram illustrating affine translation in crystals and non-affine translation in glasses.

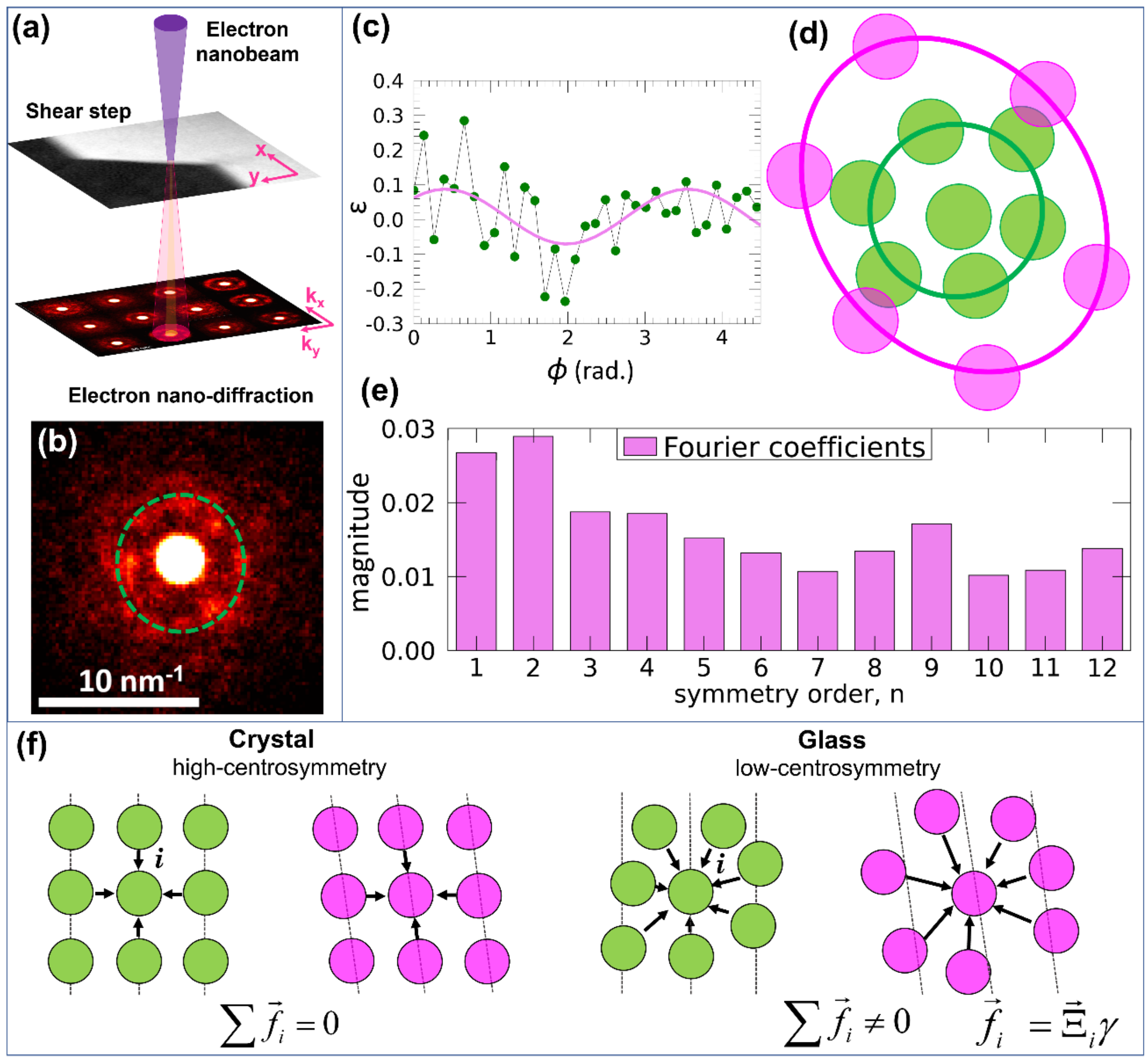

**Figure 1. Calculation of local structural parameters from scanning electron nanodiffraction (SEND)** a) The geometry of SEND in the STEM. b) Electron nanodiffraction pattern obtained from Figure a. The position of the first sharp diffraction maxima in an electron nanodiffraction pattern (b) was fitted with a strain function, $\varepsilon(\phi)$ to extract the components of the local strain ($\varepsilon_{xx}$, $\varepsilon_{yy}$) and shear strain ($\varepsilon_{xy}$) at each scan position. c) Example of a fitted strain function. d) Particle positions in the non–shear band region (green) and in the shear band (pink), highlighting the corresponding distortion of nearest-neighbour configurations. Note strains are exaggerated to highlight the changes. e) Fourier coefficients from a single END pattern revealing non-zero odd-n components, reflecting a Friedel symmetry. This results from local structures in the specimen that do not possess in-plane inversion symmetry. f) Schematic diagram illustrating the affine displacements govern the deformation of crystal and the non-affine displacements govern the deformation of glass.

## 2.2. Local Atomic Structure from Tensile Strain

To examine shear bands formed under different conditions, we bent a metallic glass ribbon so that the shiny, free-side of the ribbon (not in contact with the copper wheel during melt-spinning) underwent tension and compression. Figure 2a shows an SEM image of the tensile side, where high-contrast lines indicate the formation of a shear step under tension [55]. To study the atomic arrangement at the shear step, we prepared a cross-sectional sample of the shear band using a FIB-SEM inverted TEM lamella technique and low-temperature, inert ion polishing (see Experimental Section). This approach creates a thin area with thickness corresponding to 20-25 polyhedra while preserving the intrinsic atomic configuration beneath the shear step, enabling direct insight into the local structural arrangement below the surface.

A STEM-HAADF image of an inverted TEM lamella taken from the shear band region of a deformed $Zr_{36}Cu_{64}$ metallic glass ribbon is shown in Figure 2b. HAADF images show no clear contrast variation under the shear step, thereby limiting its effectiveness for resolving local structural heterogeneity. The lamella displays distinct layers, including a platinum (Pt) coating, a carbon (C) layer, and the underlying deformed $Zr_{36}Cu_{64}$ metallic glass. The expected location of the shear step, indicated by a yellow square, was determined using two fiducial markers placed on either side of the step. The prominence of the shear step, evident in Figures 2b and c, renders these markers as redundant after final polishing. Figure 2c presents a magnified HAADF image of this region, revealing a shear step measuring 41 *nm* in width and 64 *nm* in height. The area far from the shear step, outlined by the green square in Figure 2c, serves as the non-shear band reference region (where the reference strain $k_0$ was measured).

To investigate local atomic structural variations of deformed metallic glass, we implemented SEND with a 1.2 Å step size to simultaneously map the local degree of centrosymmetry and strain. Deformation alters both the spatial distribution and magnitude of these parameters, as seen in Figure 2d and 2e. These maps compare the non-shear band reference region, far from the shear step (green square, Figure 2c) with the shear band region beneath the shear step (pink square, Figure 2c), following the application of a tensile force perpendicular to the specimen plane. High-magnification HAADF images corresponding to the regions shown in Figure S1 (Supporting Information).

Figure 2d presents spatial maps of the degree of centrosymmetry and in-plane strain components along $x$ and $y$ directions ($\varepsilon_{xx}$ and $\varepsilon_{yy}$) and shear strain ($\varepsilon_{xy}$), obtained from the region outlined by the green square in Figure 2c as the non-shear band reference region, far from the shear step. The field-of-view of all the parameter maps is 16 nm. In the $\varepsilon_{xx}$, $\varepsilon_{yy}$, and $\varepsilon_{xy}$ maps, the color scale is centred at zero, with negative values indicated in black to green, near-zero strains in blue, and positive values ranging from pink to white. All of these structural strain maps are scaled to contrast limits of ±6 standard deviations from the mean value. Both the centrosymmetry and strain maps exhibit significant heterogeneity at the length scale of individual polyhedra, with all mapped parameters displaying complex, random, meandering, snaking and filament-like morphology.

Figure 2e shows high-resolution maps of the degree of centrosymmetry and the strain components ($\varepsilon_{xx}$, $\varepsilon_{yy}$, $\varepsilon_{xy}$) obtained from the shear band region outlined by the pink square in Figure 2c. These maps also reveal strong local spatial heterogeneity and irregularity across all parameters at the polyhedral scale. In this area just below the shear step larger spatial correlations emerge. For example, the in-plane strain component maps ($\varepsilon_{xx}$ and $\varepsilon_{yy}$) exhibit strong bands with positive strain values, with $\varepsilon_{yy}$ increasing particularly within the shear band region, highlighted by the yellow dashed lines. These bands become more visible when a low-pass Butterworth filter is applied to suppress the fluctuations at the polyhedral scale (see Figures S4, Supporting Information).

The normal strain component showing local dilation was calculated as the average of the $x$- and $y$-components $\varepsilon_n = (\varepsilon_{xx} + \varepsilon_{yy})/2$. The normal strain maps in Figure S2 (Supporting Information) also confirm strong bands with positive strain values or local dilation. The bands which we associate with the primary shear band are oriented at $\approx -45^{\circ}$ to the applied strain direction, as marked by the yellow dashed line. These patterns indicate strain localization along preferred orientations, consistent with the anisotropic deformation behavior of metallic glasses that has been observed with simulation [56]. These findings align with the simulation results presented in the subsequent section, which further support the observed strain behavior.

The shear strain ($\varepsilon_{xy}$) map shows significant variations that are consistent with the bands identified in the maps of the $x$- and $y$- strain components. In these bands, $\varepsilon_{xy}$ decreases markedly, indicating localized shear strain accumulation that has accompanied the increase in dilation (normal strain). The centrosymmetry map also displays a slight decrease in the regions corresponding to these bands, although this is slightly masked by fluctuations at the polyhedral scale. Our measurements show that a significant number of polyhedra in these zones underwent the same kind of local structural transformation to positive normal strain, negative shear strain, and decreased centrosymmetry. This suggests preferred deformation-induced structural transformation along certain pathways. The stripe-like features in the maps give rise to strongly directional features with increased correlation lengths in the 2D autocorrelation functions (Figure S4a S4b - Supporting Information). These shear bands highlight the collective response of the amorphous structure to external loading, potentially mediated by mechanisms such as shear transformation zones (STZs) or other preferred local structural transformation processes. Overall, these observations demonstrate the complexity of nanoscale strain localization and underscore the utility of high-resolution mapping in revealing subtle deformation mechanisms in metallic glasses.

Principal strain refers to the normal strains along the directions where shear strain is zero. Local strain measurements in metallic glasses represent the maximum and minimum local extension or compression at a point. The maximum and minimum principal strains ($\varepsilon_1$ and $\varepsilon_2$), and maximum shear strain ($\varepsilon_{max}$) for the two regions are presented in Figure S3 (Supporting Information). The shear bands and the extended correlation lengths are most clearly observed in the principal strain maps presented in Figure S3b (Supporting Information). The principal strain maps often show bands oriented $\approx -45^{\circ}$ to the loading direction (Figure S3b, Supporting Information). The maximum shear strain ($\varepsilon_{\max}$) shows shear strain concentrated in identified shear band zones, exhibiting a well-defined preferred orientation ($\theta_{\max} \approx -45^{\circ}$ to the applied strain direction).

The radial averages of the autocorrelation functions pertaining to each map are plotted in Figure 2f, with the width of the central maxima providing an estimate of correlated domain sizes. These autocorrelation functions show that in the regions far from the shear band high spatial frequency features at the polyhedral length scale dominate, while the region under the shear step develops broad tails indicating larger scale spatial correlations. We quantify these two length scales by fitting the central maximum of the 2D autocorrelation with two elliptical Lorentzians and report the major and minor FWHM of these (Table S2, Supporting Information). The FWHM averaged from the strain maps show a quantitative increase in the correlation length of the broader Lorentzian underneath the shear step compared to the region far away. The FWHM is perhaps an underestimation, and a more appropriate measure might be the full-width at 1/10 maximum which for Lorentzian functions is 3*FWHM. In this case, the increased spatial correlation goes from ≈2 *nm* to ≈4 *nm*, or from 3 polyhedra to 6 polyhedra (Table S2, Supporting Information). This gives us an estimate of shear band width of 4 *nm*.

Histograms of the mapped parameters reveal distinct structural populations and trends in response to tensile deformation. In the region beneath the shear step, the degree of local centrosymmetry increases slightly, and the distribution of centrosymmetry values slightly broadens (Figure 2g), indicating the coexistence of regions with varying structural order. A significant variation in local centrosymmetry between shear band and non-shear band regions is observed. This difference is probably due to a difference in thickness between the two regions. The area under the shear step is estimated to be ≈10 *nm* thicker (based on the HAADF contrast). The measure of local centrosymmetry is very sensitive to thickness as it relies on dynamical diffraction effects and so it is difficult to compare this parameter between regions with different thickness. Strain components along $x$ and $y$ directions ($\varepsilon_{xx}$, $\varepsilon_{yy}$, Figure 2g), and normal strain ($\varepsilon_n$, Figure S3, Supporting Information) show an overall positive shift, consistent with volumetric expansion, while shear strain ($\varepsilon_{xy}$) shows a negative shift, reflecting local shear accommodation. The negative sign of the shear strain indicates the shear direction, not a reduction in magnitude. The affected regions correspond to a $\approx -45^{\circ}$ shear band, suggesting that the shear strain is oriented along this direction.

To further clarify the distinction between shear band and non-shear band regions beneath the shear step and remove any thickness dependence, we selected representative regions of interest (ROIs) from each case to construct histograms (Figure S6 and S7, Supporting Information). The average and standard error of values for each structural parameter corresponding to non-shear band and shear band regions shown in Figure S6, are presented in Table 1. Clearly, shear band zones have many polyhedra that have transformed to lower centrosymmetry lower shear strain, and higher normal strain structures. The reported strain changes are derived from the mean values in Table 1, calculated from the ROIs defined in Figures S6 (tension) and S12 (compression) beneath the shear step. These ROIs correspond specifically to shear band and adjacent non-shear band regions. In addition, the normal strain ($\varepsilon_n$) increases from 0.0297 in the non-shear band region to 0.0371 within the shear band, corresponding to a 24.9% increase. In contrast, the shear strain becomes more negative, changing from −0.00925 to −0.0297, a 221% decrease. The mean values and standard errors for the structural parameters corresponding to the regions shown in Figure S7 are summarized in Table S1 and exhibit the same trends. These trends show how deformation causes local structures to dilate (free volume creation) and at the same time undergo shear distortions, that also correlate to lower structural centrosymmetry. We discuss these trends further after examining the case of deformation in compression. Deformation in tension produces excess free volume or atomic rearrangement that led to local expansion [57]. The HAADF image from the shear band region displays uniform contrast, suggesting variations in free volume are correspondingly subtle. Our strain maps derived from nanodiffraction show that this free volume creation occurs in bands and is small, at the level of 0.04%, information that is difficult to extract from the HAADF images.

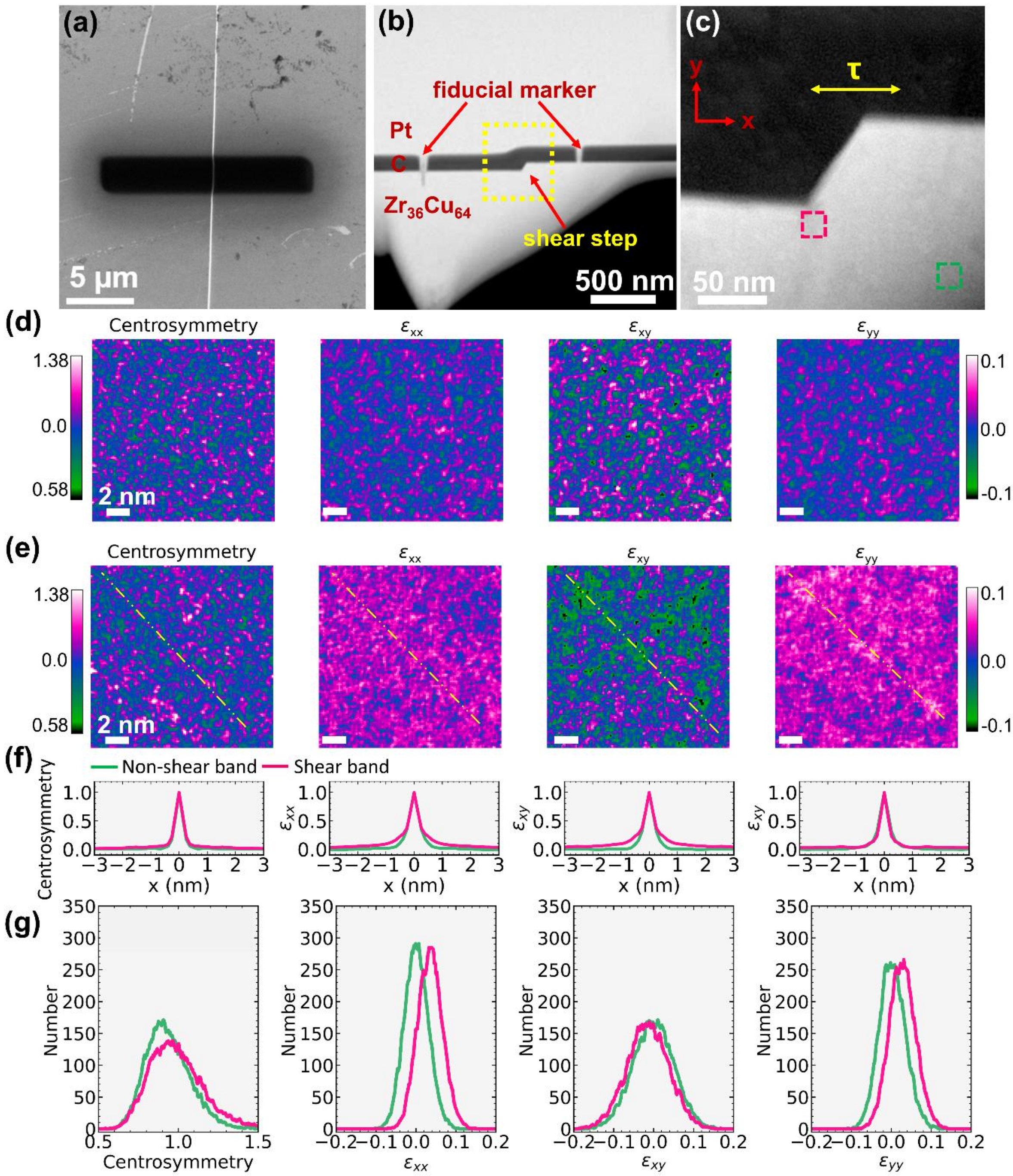

**Figure 2. Local atomic rearrangements under tensile strain in deformed** $Zr_{36}Cu_{64}$ **metallic glass ribbons.** a) SEM image showing a shear step formed by a shear band in the deformed metallic glass. The black carbon layer indicates the location of the inverted TEM lamella. b) STEM-HAADF image of the inverted TEM lamella; the yellow square marks the shear band location, determined from two fiducial markers on either side of the shear step. c) Higher-magnification STEM-HAADF image of the shear band, showing the regions where SEND maps were acquired (green and pink squares). d) Degree of local centrosymmetry and the strain components ($\varepsilon_{xx}$, $\varepsilon_{yy}$, $\varepsilon_{xy}$) within the region indicated by the green square. The color scale for local centrosymmetry is on the left, and the color scale for strain maps is on the right. e) The degree of local centrosymmetry map and strain maps in the area indicated by the pink square. f) Radial averages of the autocorrelation function for non-shear band and shear band regions. g) Histograms of mapped parameters.

## 2.3. Local Atomic Structure from Compressive Strain

Figure 3a shows an SEM image of the compression side surface of a metallic glass ribbon where high contrast lines indicate the formation of shear step under compression. A STEM-HAADF image of the inverted TEM lamella in Figure 3b clearly reveals distinct layers within the sample: a Pt layer, a C layer, and the deformed $Zr_{36}Cu_{64}$ metallic glass subjected to compressive stress. A yellow square highlights the anticipated location of the shear step, confirmed using two fiducial markers placed on the sample surface. Figure 3c provides a magnified HAADF image of the region marked in Figure 3b, clearly revealing the shear step, which measures approximately 22 *nm* in width and 38 *nm* in thickness.

High magnification HAADF images of shear band and non-shear band regions are shown in Figure S8a and b, respectively (Supporting Information). As discussed above, the shear band is not clear from contrast variations in the HAADF image. The area far from the shear step, outlined by the green square in Figure 3c, serves as the non-shear band reference region. Figure 3d shows high-resolution maps of the degree of centrosymmetry along with the in-plane components of the strain ($\varepsilon_{xx}$, $\varepsilon_{yy}$, $\varepsilon_{xy}$), obtained from the non-shear band region outlined by the green square in Figure 3c. The centrosymmetry and strain distributions are highly heterogeneous, with the observed parameters exhibiting complex, non-uniform, and filament-like arrangements.

Figure 3e shows the strain components along $x$ and $y$ directions ($\varepsilon_{xx}$, $\varepsilon_{yy}$) of the shear-band region beneath the shear step. As for the region under the shear step in tension, the parameter maps reveal bands 3−5 *nm* in width. The yellow dashed line marks the primary shear band, oriented at ≈45° to the applied strain direction. These primary shear bands are also observed in normal strain maps in Figure S9b (Supporting Information) and enhanced by a low-pass filter that suppresses fluctuations at the polyhedral length scale (Figure S11c Supporting Information). Concurrently, the local degree of centrosymmetry and strain maps develop two distinct correlation lengths: a short, fluctuating one at the polyhedral size, and a longer-range one, indicative of cooperative transformations within the bands. These trends are evident in the radial averages of the autocorrelation function for each map are presented in Figure 3f and also in the 2D autocorrelation functions that show strong directional streaking (Figure S11a, b Supporting Information). We further quantify these tends by fitting two elliptical Lorenztians to the central maximum of the 2D autocorrelation function. The broader Lorentzian full-width at 1/10 maximum increases from ≈2 *nm* to ≈3 *nm*, or from 2 polyhedra to 4 polyhedra (Table S2, Supporting Information). This analysis does not model the larger scale, diffuse and directional intensity in the 2D autocorrelation function but with larger area spatial imaging and better statistics, this could be done to compare to previous measurements of softness and structural variation in and around shear bands, which distinguish a localized core from broader, more diffuse surrounding zones [58,59].

These coordinated structural changes are difficult to observe in the correlation lengths of the radially averaged autocorrelation functions, as the changes are subtle.

Relative to a region far from the shear step, the degree of centrosymmetry in local structures has only slightly decreased, while the width of this distribution appears unchanged (Figure 3f). In comparison, the distribution of $\varepsilon_{xx}$ and $\varepsilon_{yy}$ show minimal broadening. The most prominent observation is that the distributions are shifted to the right, indicating a significant overall increase. The average values of $\varepsilon_{xx}$ and $\varepsilon_{yy}$ and normal strain ($\varepsilon_n$) has increased, indicating an overall generation of volume. This may seem counterintuitive, but net dilation caused by local structural transformations in disordered media is well known, even if the nature of the deformation is compressive [21,56,60]. The local degree of centrosymmetry exhibits only minimal variation, even with a significant increase in normal strain. The maps suggest that this behaviour arises because the shear band regions undergoing symmetry-lowering transformations are not isolated but become spatially connected. As these linked regions develop, they form continuous bands that concentrate strain, leading to low-stability zones with high normal strain. This spatial organisation provides a key explanation for why the centrosymmetry distribution remains relatively uniform while strain becomes highly localized. Other simulation studies have also demonstrated that structural softening along the pathway to failure only requires a remarkably slight reduction in local centrosymmetry [47]. The results are consistent in both the specimens deformed under tensile and compressive strains.

The map of centrosymmetry, overlaid or shown alongside the strain components, further reinforces the notion of strong nanoscale disorder. Regions of reduced centrosymmetry correspond well with areas of elevated strain, implying that local structure distortions directly perturb the local symmetry environment. This correlation is consistent with scenarios in which strain gradients couple to symmetry breaking mechanisms, such as local phase transitions, polar distortions, discontinuous yielding, boson peak, or flexoelectric effects in centrosymmetric materials [48,61].

The maximum and minimum principal strains ($\varepsilon_1$ and $\varepsilon_2$), and maximum shear strain ($\varepsilon_{max}$) for the non-shear band and shear band regions are presented in Figure S10 (Supporting Information). Figure S10b indicates maximum shear strain ($\varepsilon_{\mathrm{max}}$) in the shear-band region beneath the shear step, where it is concentrated in the identified shear band zones. These bands typically align at a well-defined orientation $\theta_{max} \approx 45^{\circ}$, relative to the loading axis, which is consistent with the theoretical plane of maximum shear stress predicted in previous studies under simple loading conditions [62]. The marked dilation observed suggests that while there is some volume change, the structural transformation is primarily driven by shear deformation, as indicated by the ≈45° orientation of the shear bands. In metallic glasses, shear bands oriented at ≈45° indicate that structural transformation occurs through highly localized, shear driven atomic rearrangements. These bands serve as regions of intense plastic flow and potential softening, while the surrounding material largely retains its elastic behavior.

Representative ROIs were selected to compare shear and non-shear band regions beneath the shear step (Figure S12 and S13, Supporting Information). The average and standard error of values for each structural parameter corresponding to the shear band and non-shear band regions shown in Figure S12, are displayed in Table 1, confirming the trends previously noted of positive strain along *x* and *y* directions and negative shear strain in the regions identified as narrow shear bands. Quantitatively, $\varepsilon_n$ increases from 0.01995 to 0.0266 (a 33.3% increase), whereas the $\varepsilon_{x\gamma}$ decreases from -0.0165 to -0.0265, corresponding to a 60.6% decrease. The mean values and standard errors for the structural parameters corresponding to the regions shown in Figure S13 are summarized in Table S1 and exhibit the same trends.

These observations illustrate the highly localized and non-uniform nature of plastic deformation in amorphous materials, where factors such as shear band dilation, asymmetric yielding, shear steps, and atomic rearrangements within STZs can produce local volume expansion [63,64]. Primary shear band typically form along preferred orientations (≈45° to the loading axis) based on internal structural heterogeneity and local stress concentrations rather than the global bending direction, leading to a behavior that appears symmetric for upward and downward bending.

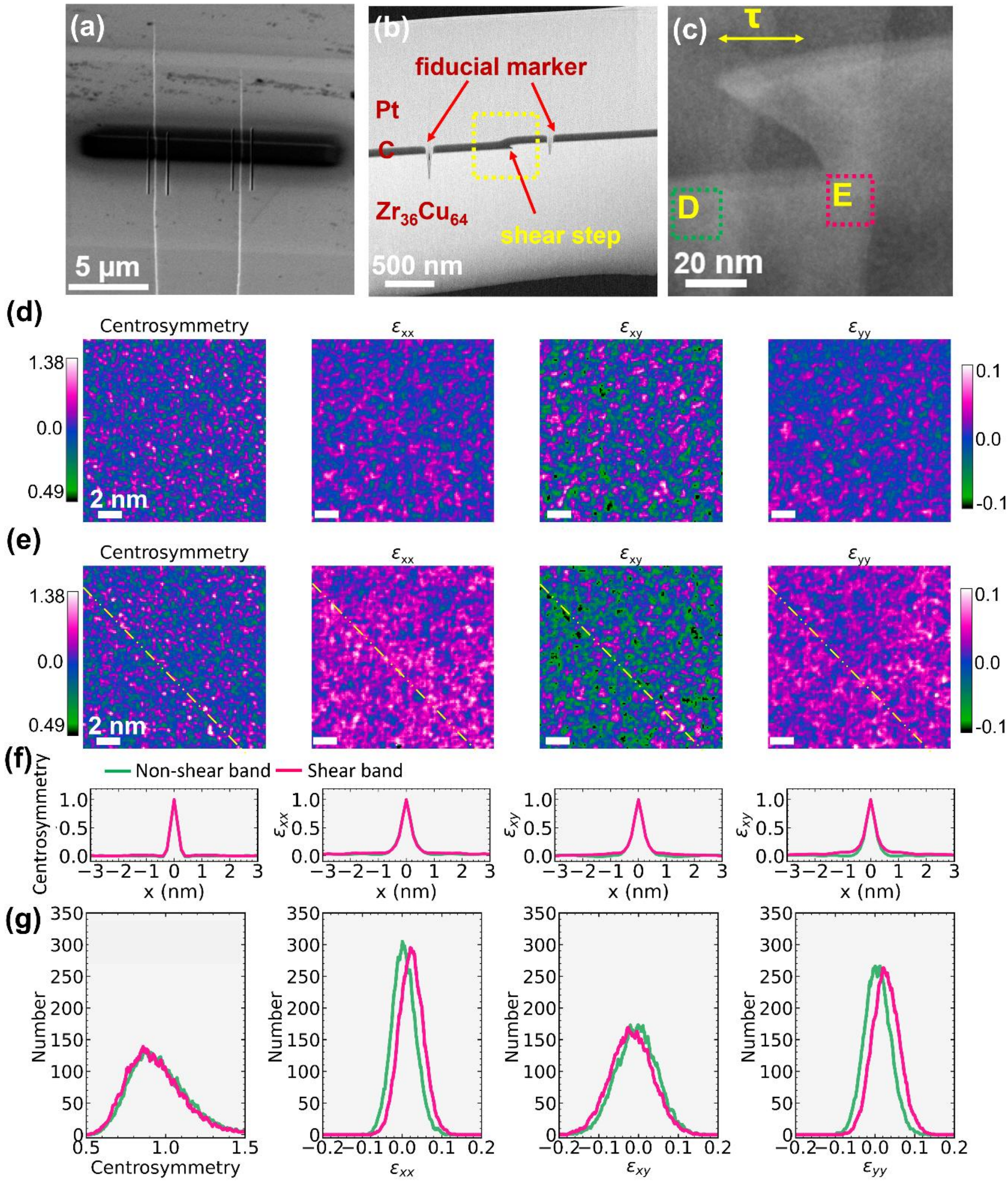

**Figure 3. Local atomic rearrangements under compressive strain in deformed** $Zr_{36}Cu_{64}$ **metallic glass ribbons.** a) SEM image showing a shear step formed by a shear band in the deformed metallic glass. The black carbon layer indicates the location of the inverted TEM lamella. b) STEM-HAADF image of the inverted TEM lamella; the yellow square marks the shear band location, determined from two fiducial markers on either side of the shear step. c) Higher magnification STEM-HAADF image of the shear band, showing the regions where SEND maps were acquired (green and pink squares). d) The degree of local centrosymmetry and in-plane strain components ($\varepsilon_{xx}$, $\varepsilon_{yy}$, $\varepsilon_{xy}$) in the non-shear band area indicated by the green square in Figure c. The color scale for local centrosymmetry is on the left, and the color scale for strain maps is on the right. e) The degree of local centrosymmetry map, in-plane strain components and shear strain maps in the shear band area indicated by the pink square in Figure c. f) Radial averages of the autocorrelation function for non-shear band and shear band regions. g) Histograms of mapped parameters.

### 2.4. Simulated Local Atomic Structure from Linear Tensile Strain

To validate our observations, we employed the multislice method [65] to simulate electron nanodiffraction from part of a large (49 × 245 × 490 Å$^3$) $Zr_{50}Cu_{50}$ atomic model (shown in Figure 4a) from molecular dynamics strained in tension to $\varepsilon = 0.2$ in the $x$-direction [56]. In brief, the multislice method realistically captures dynamical electron diffraction effects by calculating the transmission function of the electron wave from each slice of the atomic model and propagating the wave between slices [65]. We then applied the same analysis to extract local centrosymmetry and strain from the simulated diffraction patterns. The displacement maps calculated from the actual atomic displacements reveal the anisotropic and spatially heterogeneous deformation that gives rise to the formation of a large primary shear band in this area (Figure 4b-c). In particular, the difference in $x$-displacements from positive in the bottom left to negative in the top right (Figure 4b) reveals the position where the strain in the $x$-direction is localized in the primary shear band ≈45° to the applied strain direction (Figure S14, Supporting Information).

This primary band was identified in the original work from a large increase in von Mise strain [53]. Figure 4d shows maps of local centrosymmetry alongside the in-plane strain components ($\varepsilon_{xx}$, $\varepsilon_{yy}$, $\varepsilon_{xy}$) for the region highlighted in Figure 4b. The maps reveal shear band formation oriented approximately 45° to the loading axis.

Representative ROIs were chosen to compare shear and non-shear band regions (Figure 4c). ROI1 (orange), taken from the surrounding matrix, represents the non-shear band area, whereas ROI2 (blue) corresponds to the shear band area. Although the overall degree of centrosymmetry is comparable between the two regions, it exhibits a slight shift toward lower symmetry in the shear band, accompanied by distinct differences in strain behavior. Specifically, the strain components along $x$ and $y$ directions ($\varepsilon_{xx}$ and $\varepsilon_{yy}$) shift toward positive values within the shear band, while the shear strain ($\varepsilon_{xy}$) becomes more negative (Figure 4d).

Histograms of the mapped parameters are shown in Figure 4e. The direction and magnitudes of these observations are consistent with the experimental results described in the previous section for tensile and compressive strain. The average and standard error of values for each structural parameter corresponding to shear band and non-shear band regions are displayed in Table 1. The change in centrosymmetry observed in the diffraction simulation is consistent with both the compression and tensile experiment results. In all cases, the centrosymmetry decreases in the shear band region relative to the non-shear band region, indicating increased local structural disorder. Quantitatively, the reduction is ~1.6% under tension, ~1.89% under compression, and ~2.57% in the simulation, showing comparable magnitudes.

The simulations employ idealized structural models and simplified loading conditions that may not fully capture the structural heterogeneity and experimental environments of real $Zr_xCu_{100-x}$ metallic glasses. In addition, finite model size and limitations associated with the interatomic potentials may influence the quantitative agreement with experiments. Nevertheless, the simulations successfully reproduce the key qualitative trends and underlying deformation mechanisms observed experimentally.

**Table 1.** Average values and standard errors of the mean for structural parameters from the shear band and non-shear band regions, obtained from ROIs in Figure S4 (tension), Figure S9 (compression), and Figure 4 (simulation).

| Regions | Measurements | Tension | Compression | Simulation |
|---|---|---|---|---|
| Non-shear band | Centrosymmetry | 0.995 ± 0.0042 | 0.950 ± 0.0047 | 2.33 ± 0.024 |
| | $\varepsilon_{xx}$ | 0.0329 ± 0.0008 | 0.0186 ± 0.0007 | 0.0204 ± 0.00055 |
| | $\varepsilon_{xy}$ | −0.00925 ± 0.0013 | −0.0165 ± 0.0013 | −0.00277 ± 0.00079 |
| | $\varepsilon_{yy}$ | 0.0265 ± 0.0008 | 0.0213 ± 0.0008 | 0.000298 ± 0.00005 |
| Shear band | Centrosymmetry | 0.979 ± 0.0042 | 0.9320 ± 0.0048 | 2.27 ± 0.021 |
| | $\varepsilon_{xx}$ | 0.0430 ± 0.0008 | 0.0209 ± 0.0007 | 0.297 ± 0.00069 |
| | $\varepsilon_{xy}$ | −0.0297 ± 0.0013 | −0.0265 ± 0.0013 | −0.0103 ± 0.0009 |
| | $\varepsilon_{yy}$ | 0.0312 ± 0.0008 | 0.0323 ± 0.0008 | 0.000453 ± 0.00005 |

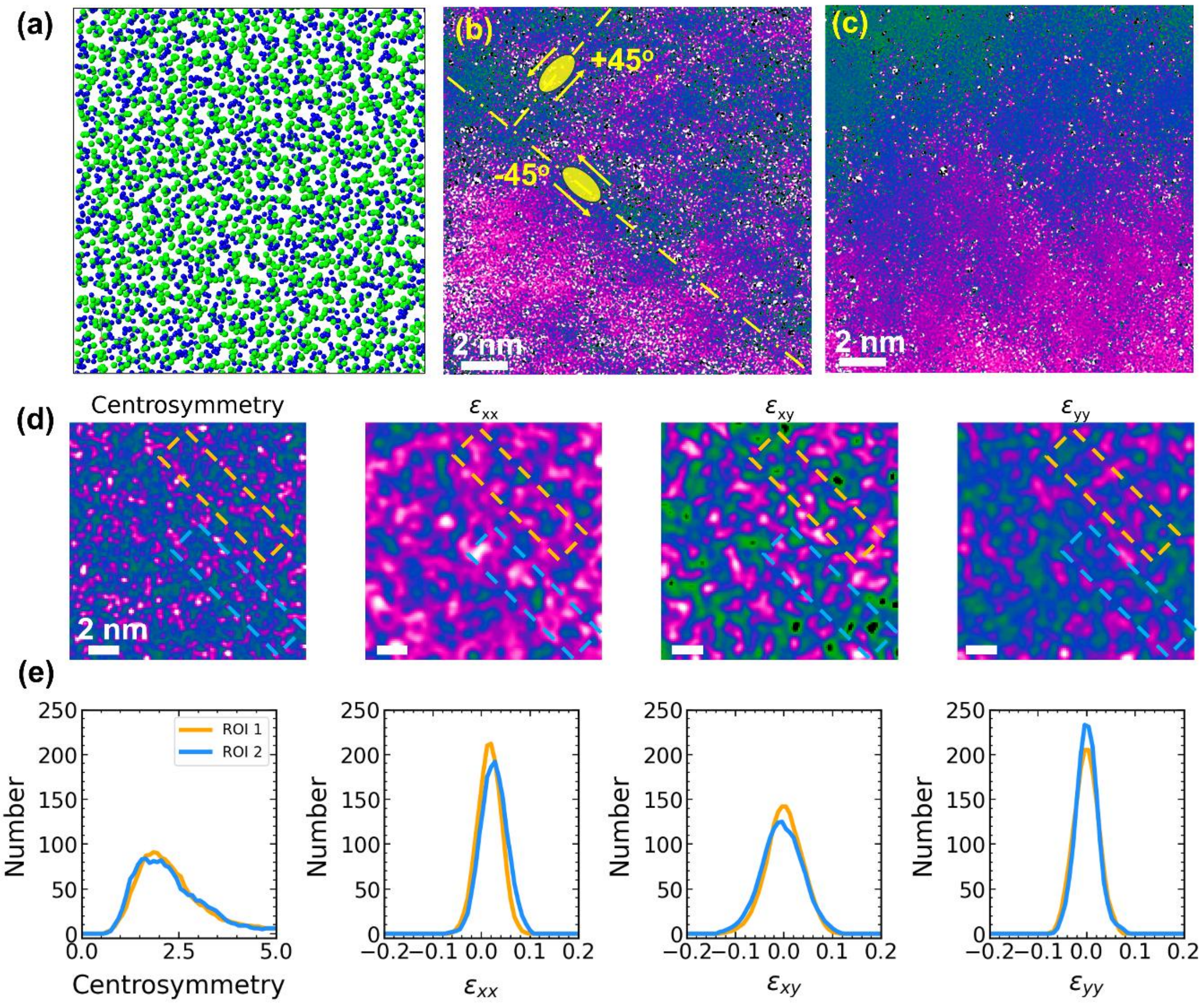

**Figure 4. Simulated local atomic rearrangements in deformed metallic glass ribbons.** a) Atomic structure model of metallic glass. Green atoms are Zr, blue atoms are Cu. b) Displacement field along *x* axis of the metallic glass. c) Displacement field along *y* axis of the metallic glass. d) Simulated degree of local centrosymmetry map and in-plane strain component maps ($\varepsilon_{xx}$, $\varepsilon_{yy}$, $\varepsilon_{xy}$) of area in Figure b. The non-shear band region, indicated by the orange square (ROI1), was chosen from the surrounding matrix to serve as a reference, whereas the shear band region, indicated by the blue square (ROI2), was selected to capture the structural response within the highly deformed zone. e) Histograms of mapped parameters from non-shear band and shear band regions.

## 3. Discussion and Conclusion

We report a novel method for preparing inverted TEM lamellae of shear band cross-sections in deformed metallic glasses that can preserve the structure of the shear band underneath a shear step for *post facto* analysis by sensitive electron nanodiffraction methods. By combining precise fiducial markers, inverted FIB-SEM milling, and low-temperature ion polishing, we generated high-quality specimens. The inverted lamellae and low-temperature polishing preserve the subtle structural differences between shear band regions where strain has localized underneath shear steps and non-shear band regions. This allowed direct, artifact-free analysis of very small (~1–2%) structural transformations, quantified here as changes in the centrosymmetry parameter between non-shear band and shear band regions.

The uniform contrast observed in HAADF imaging even within highly deformed regions, limits our ability to characterise the local atomic structure of metallic glass using imaging alone. However, shear bands are clearly revealed by more sensitive, scanning diffraction-based techniques.

By combining a fast electron detector with electron nanodiffraction, we can probe local atomic structures and nanoscale strain beyond the reach of conventional imaging. A finely focused beam and controlled specimen thickness produce speckle diffraction patterns that encode rich information about local symmetry and atomic structure. By analysing the first speckle ring, we quantified nanoscale strain and mapped the in-plane strain components in deformed $Zr_{36}Cu_{64}$ metallic glass. Additionally, deviations from Friedel symmetry in the speckle pattern enabled measurement of local centrosymmetry. We correlate deformation-induced local structural anisotropy (strain) with local centrosymmetry in $Zr_{36}Cu_{64}$ metallic glass. Together, these capabilities show that speckle diffraction provides a uniquely powerful window into the hidden structural and symmetry variations in disordered materials such as glasses.

Overall, the changes in the structural parameters we measure are extremely subtle in our *post facto* analysis which could in part be due to the relaxation of stress built up during deformation in the metallic glass, resulting in weaker residual features. Our experimental results (both the trends and their magnitude) are in excellent agreement with atomistic simulations and prior diffraction-based measurements on colloidal glasses, reinforcing the validity and universality of the observed phenomena [46,47].

Spatially resolved measurements of these sensitive structural parameters reveal inhomogeneity at the length scale of individual polyhedra in agreement with previous investigations of shear banding in metallic glasses [21]. Our high-spatial resolution measurements are also consistent with the length scale for heterogeneous response from anisotropies in x-ray diffraction, but unlike this bulk measurement, we map these heterogeneities directly and at the nanoscale and moreover, the local centrosymmetry parameter has a strong connection to theory [48–50,66].

Despite the fact that all structural parameters exhibit broad distributions, we see significant, marked changes in the most deformed regions underneath shear steps. In these regions, spatial correlations increase (larger-scale inhomogeneity), due to the formation of ~3 *nm* thick shear bands where the normal strain increases by 24.9 % and the shear strain decreases by 221 % under compression. In tension, the normal strain increases by 33 %, while the shear strain decreases by 60.6 %. This strain localization is a key mechanism underlying discontinuous yielding in metallic glass materials [61,67,68].

These large changes in local strain in shear bands are accompanied by a small decrease in local centrosymmetry of ≈1–2%. This decrease is significantly larger than the uncertainties, has been observed previously and is also consistent with the simulations we perform in this work from a molecular dynamics model and previous experiments on colloidal glasses [46,47]. A decrease in the degree of centrosymmetry causes unbalanced affine forces that drive extra non-affine displacements of the atoms or rigid units [69], producing a negative correction to the shear modulus [49,70] and, thus, an overall mechanical softening of the material (Figure 1f). Hence, even a tiny decrease in the local centrosymmetry, possibly coupled to a local decrease in atomic connectivity, can induce a significant drop in the local rigidity, which in turn leads to dramatic yielding via self-organization of soft spots into shear bands. This pattern in local structural parameters (decreased centrosymmetry, increased strain away from reference values) seems to be a hallmark of the coordinated local structural transformations after large scale slip in shear bands.

In our strain maps these shear-bands form subtle but distinct striped patterns, with primary shear bands consistently emerging at ≈45° to the loading axis under both tension and compression, agreeing with simulation predictions [56,62,71,72]. In $Zr_xCu_{100-x}$-based metallic glasses, shear bands consistently form at this angle, aligned with the direction of maximum shear strain, regardless of whether the material is under tension, compression, or combined loading conditions [71]. This behavior is consistent with the maximum shear stress theory and is observed in both tension and compression [62,71]. Additionally, softening within the shear bands further reduces resistance to deformation, ensuring that strain remains localized and comparable under both stress states [62,72].

The large increase in local normal strain, or volume dilation was observed in both shear bands formed under tension and compression. This creation of free volume due to deformation is widely reported [73,74]. Our measurements show that shear bands generate complex local strain states characterized by volumetric changes and significant shear distortion. Specifically, the shear bands produce large shear strains as atoms undergo coordinated slips along preferred planes. Previous simulation studies have shown complex networks of shear bands in metallic glasses that are oriented at both −45° and +45° to the loading axis [56]. In our shear bands that are oriented at −45°, we observe a significant decrease in shear strain which, in our coordinate frame, we attribute to the structural extension of the polyhedra along the shear band axis.

Accordingly, these local observations warrant further investigation over larger fields of view. However, our experiments and simulations have demonstrated the necessity of spatial sampling below the length scale of individual coordination polyhedra. While anticipated to generate prohibitively large data sets, such studies could unveil the connection between these heterogeneous local transformations with the formation of complex shear band networks in plastically deformed soft zones. Future work should combine larger-area sampling with the same sensitivity to local structure in combination with *in situ* techniques (e.g., SEM and STEM) to better capture structural evolution during deformation.

Our sensitive, nanoscale observations provide crucial insight into the deformation mechanisms of metallic glasses, demonstrating how localized atomic rearrangements collectively shape their macroscopic mechanical response. We also demonstrate a unified methodology capable of directly imaging multiscale phenomena in metallic glass, thereby bridging the critical gap between nanoscale structure, multiscale deformation processes, and bulk material response. Importantly, the results show that these atomic-scale strain and symmetry correlations, first established in colloidal glasses, also reveal key aspects of the complex behaviour of strained metallic glasses. Multislice simulations of scanned diffraction patterns from atomic models further confirm these findings. Together, they underscore the broad applicability and robustness of our approach. These sensitive local structural parameters provide a new experimental approach to test microscopic theories of deformation in glasses and could be employed to study other phenomena in different amorphous systems.

## 4. Experimental Section

### 4.1. Inverted TEM lamella

We examined a melt-spun $Zr_{36}Cu_{64}$ metallic glass ribbon prepared under argon (oxygen level 15 ppm) with a copper wheel speed of 25 m/s and a melt-temperature of 1170$^{\circ}$ C. Shear bands were formed on the shiny, free-sides of the ribbon in both compression and tension by bending the ribbons at room temperature. This process produced shear steps appearing as high-contrast lines in the SEM image (Figure S15a, Supporting Information).

Glasses are not in equilibrium, and TEM preparation methods using energetic ions often cause structural modifications that are difficult to detect in a disordered structure. Additionally, shear band areas are often challenging to identify in a *post facto* analysis due to very small structural changes that are not evident as a contrast difference in typical imaging methods. The inverted lamella preparation method is essential for preserving the intrinsic atomic structure beneath the shear step because it minimizes ion beam induced damage in the most structurally sensitive region.

To prepare thin (ideally 20 – 30 *nm*), clean, cross-sectional TEM specimens of the regions just underneath the shear step we employed a combination of FIB-SEM and precision ion polishing system (PIPS) to prepare an inverted TEM lamella.

Site specific cross-section preparation was performed using a Thermo Fisher Helios 5 UX. A 1 $\mu$m protective carbon (C) layer was first deposited over the shear band using a low-energy electron beam (2 kV, 13 nA) (Figure S15b, Supporting Information). Next, two fiducial markers were milled (30 kV and 7 pA) 600 *nm* from either side of the shear band to identify the shear band if shear step is small (Figure S15b, Supporting Information). A second 1 $\mu m$ platinum (Pt) protective layer was then deposited on top of the carbon layer further protect the metallic glass structure at the surface during subsequent processing (Figure S15c, Supporting Information).

With the protective layers in place, we began rough milling the surrounding material using a $Ga^+$ ion beam to define the lamella (Figure S15c, Supporting Information). This was followed by the application of a J-cut technique (Figure S15d, Supporting Information), in which the lamella was cut from the bottom and one side to allow for clean extraction. After milling, both sides of the lamella were cleaned using a cross-section cleaning process (Figure S15e, Supporting Information). The lamella was then rotated 90$^\circ$ (Figure S15f, Supporting Information) before being lifted from the bulk material using a micromanipulator (Figure S15g-h, Supporting Information). Subsequently, the lamella was attached to a molybdenum (Mo) TEM half-grid (Figure S15i, j, Supporting Information). This inversion technique positioned the platinum-coated surface at the bottom (Figure S15k, Supporting Information), which is critical for minimizing platinum re-deposition down the lamella surface during final thinning.

The inverted lifted-out lamella was thinned down to approximately 50 *nm* using a series of progressively lower-energy ion beam steps: 5 kV at 63 pA, 2 kV at 44 pA, and finally 0.5 kV at 50 pA. This gradual reduction in ion energy helped to produce a smooth, uniform surface optimized for high-resolution TEM imaging (Figure S15l, Supporting Information). The use of the inversion technique during thinning altered the sputtering geometry, effectively minimizing platinum (Pt) implantation or re-deposition into the region of interest and preserving the integrity of the underlying shear band structure.

To prevent surface oxidation, the prepared lamella was transferred under vacuum in a nitrogen-filled suitcase to a Gatan 695 PIPS II for final polishing. Here, knock-on damage and local heating during final polishing were minimised by operating at low temperature (−150 $^\circ$C) and using a gentle ion beam, typically Argon ($Ar^+$), (0.5–2 keV and ±2$^\circ$ glancing angle). This final step (typically 5-10 minutes) further thinned the specimen to approximately 20 – 30 *nm*, removed any residual surface contamination, and preserved the intrinsic atomic structure beneath the shear step (Figure S16, Supporting Information). The final polished lamellas were inserted into the S/TEM within 20 minutes of final polishing to prevent the formation of any surface oxide.

### *4.2. Method*

Speckle patterns from previous soft x-ray and electron studies reveal the breakdown of Friedel or inversion symmetry in multiple scattering or dynamical diffraction patterns from the glass $[I(k,\phi) \neq I(k,\phi+\pi)]$ [42,46]. This inversion symmetry breaking is sensitively captured by the presence of odd-order symmetries in the angular autocorrelation function of the diffraction pattern, quantified by non-zero odd Fourier coefficients ($c^{2n+1} > 0$). While this work builds upon the symmetry-based structural analysis pioneered in previous studies [42,46], which first established the structural relevance of odd-symmetry components in dynamically scattered x-ray patterns, our approach extends the utility of this symmetry metric beyond $\mu$-SAXS imaging to electrons that interact more strongly than x-rays and are thus more sensitive to structural centrosymmetry revealed by dynamical diffraction effects. To our knowledge, this study represents the first application of this Friedel symmetry-breaking SEND analysis to metallic glass.

The angular autocorrelation function ($C(k, \Delta)$) (see equation 1) [40,41,75] was used to analyze each diffraction pattern in order to detect subtle angular symmetries at different values of scattering vector magnitude. Here the diffraction intensity is represented in polar coordinates as ($I(k_x, k_y) = I(k, \phi)$).

$$C(k,\Delta) = \frac{\langle I(k,\phi)I(k,\phi+\Delta)\rangle_\phi - \langle I(k,\phi)\rangle_\phi^2}{\langle I(k,\phi)\rangle_\phi^2} \qquad (2)$$

Here the scattering vector magnitude is $|\vec{k}| = k = \sqrt{k_x^2 + k_z^2}$, $k = 2sin\theta$ with a scattering angle of $\theta$ and electron wavelength of $\lambda$. $\phi$ is the azimuthal angle in the diffraction plane and $\langle X\rangle_\phi = \frac{1}{2\pi}\int_0^{2\pi} X\,d\phi$ denotes averaging over $\phi$. Normalized symmetry magnitudes $c_k^n/c_k^0$ (Fourier coefficients) are calculated from the Fourier transform of $C(k, \Delta)$ [54] at each value of $k$. The degree of local centrosymmetry is calculated by taking the ratio of even to odd Fourier coefficients: $\sum c^{2n+2}/\sum c^{2n+1}$ for $0 \leq n \leq 5$.

To reliably estimate local strain from slight anisotropies in the speckle diffraction patterns with highly discontinuous rings of intensity the following procedure was employed. The patterns were divided into 24 angular segments and the intensity in each segment was averaged over azimuthal angle, $I_\phi(k)$. The intensity centre-of-mass (CoM) in each segment $k_{max}(\phi) = \sum_n k_n I_\phi(k_n)$ was calculated to give the position of the first diffracted maximum, $k_{max}$ in that azimuthal direction, $\phi$. $k_{max}(\phi)$ was fitted with a strain function $\varepsilon_\phi$ [17,37] to extract the components of the symmetric strain tensor in the plane of the specimen, $\varepsilon_{xx}$, $\varepsilon_{xy}$ and $\varepsilon_{yy}$. Figure 1c shows an example of a fitted strain function. This illustrates that the procedure of angular averaging to smooth noise from the discontinuous ring and CoM method to determine the position of the first diffracted ring along different azimuthal directions results in a robust fitting method.

The local normal strain $\varepsilon_n$ and shear strain $\varepsilon_{xy}$ components map distortions in polyhedra resulting from local dilation and contraction (positive and negative values of $\varepsilon_n$) and volume-conserving shear ($\varepsilon_{xy}$). Principal strain refers to the maximum and minimum normal strains that occur at a specific point in a material, oriented along specific directions where shear strain is zero [16]. These strains are oriented along particular directions, known as principal directions, where the shear strain is zero. In these directions, the material undergoes pure stretching or compression without any shear distortion. The principal strains are calculated as follows Equation 3 and 4:

$$\varepsilon_1 = \frac{1}{2}\left(\varepsilon_{xx} + \varepsilon_{yy}\right) + \sqrt{\frac{1}{4}\left(\varepsilon_{xx} - \varepsilon_{yy}\right)^2 + \varepsilon_{xy}^2} \quad (3)$$

$$\varepsilon_2 = \frac{1}{2}\left(\varepsilon_{xx} + \varepsilon_{yy}\right) - \sqrt{\frac{1}{4}\left(\varepsilon_{xx} - \varepsilon_{yy}\right)^2 + \varepsilon_{xy}^2} \quad (4)$$

here, $\varepsilon_1$ is the maximum normal strain occurring at the point, and $\varepsilon_2$ is minimum normal strain at the point.

The principal directions are the orientations along which the principal strains occur. These directions can be determined by following Equation 5:

$$\tan 2\theta = \frac{2\varepsilon_{xy}}{\varepsilon_{xx} - \varepsilon_{yy}} \quad (5)$$

where $\theta$ is the angle of principal directions relative to the $x$ axis.

The maximum shear strain represents the greatest amount of shear deformation that occurs at a specific point in the material. This maximum shear strain happens at a particular angle relative to the principal strain directions. The formula for maximum shear strain:

$$\varepsilon_{xy,max} = \frac{1}{2}(\varepsilon_1 - \varepsilon_2)) \quad (6)$$

This maximum shear strain occurs at the point where the material is undergoing the most deformation in the form of shear, which corresponds to a direction at 45° relative to the principal strain directions.

Centrosymmetry and strain maps were smoothed using a Savitzky-Golay filter [76] with a window width of 4 pixels and a polynomial order of 2. This least-squares smoothing method effectively reduces noise while preserving the original peak widths and magnitudes. Correlation lengths were extracted from the two-dimensional autocorrelation functions of raw (unfiltered) maps by fitting them with two elliptical Lorentzian functions with arbitrary rotations using the MPFIT2DFUN Levenberg-Marquardt algorithm (see Figure S5, Supporting Information) [77]. The correlation lengths reported in Table S2 are the averages from the strain maps for each specimen. To aid in the identification of shear bands from the maps of structural parameters, the raw maps were filtered using a low-pass Butterworth filter with a cutoff of 3 and an order of 3. This filter attenuated spatial frequencies above 1 $nm^{-1}$ to zero, with a soft cut-off below this, and thus enhanced low spatial frequency features and filtered out the higher spatial frequencies corresponding to variations at the polyhedral length scale. Employing this method larger scale, linear features with distinct structural signatures could be identified (see Figures S4 and S11, Supporting Information) as shear bands as indicated in Figures 2 and 3 with yellow dashed lines.

Maps of centrosymmetry and strains are derived exclusively from electron nanodiffraction patterns, offering new opportunities to identify soft zones in glasses and explore their relationship with local structures. This approach also enables direct measurement of the size and spatial distribution of soft zones, providing insights into their role in multiscale phenomena such as deformation and rejuvenation [78,79]. As a diffraction-based technique, it offers a compelling alternative to the experimental challenge of resolving particle positions in three dimensions, particularly in bulk glasses and the large volumes needed to study multiscale behavior.

**Acknowledgements**

The authors acknowledge support from the Australian Research Council (DP250102966, FT180100594). The authors acknowledge the use of the instruments and scientific and technical assistance of Dr Yang Liu at the Monash Centre for Electron Microscopy, Monash University, a Microscopy Australia (ROR:042mm0k03) facility supported by NCRIS. This research used equipment funded by Australian Research Council grants LE200100132, LE0454166).